\documentclass{PoS}

\title{Observational constraints on a decaying cosmological term}
\ShortTitle{Observational constraints on a decaying cosmological term}

\author{\speaker{Riou Nakamura}\\
   Department of Physics,  Graduate school of sciences, 
   Kyushu University, 
    Fukuoka 810-8560, Japan \\
	E-mail: \email{riou@gemini.rc.kyushu-u.ac.jp}}
	
	\author{Masa-aki Hashimoto \\
	Department of Physics, Faculty of sciences, 
	Kyushu University 
	Fukuoka 810-8560, Japan \\
	E-mail: \email{hashi@gemini.rc.kyushu-u.ac.jp}}
	
	\author{Kiyotomo Ichiki \\
	Research Center for the Early Universe, 
	Graduate School of Science, 
	The University of Tokyo
	Tokyo 113-0033, Japan \\
	E-mail: \email{ichiki@resceu.s.u-tokyo.ac.jp}}

\abstract{
 We investigate the evolution of a universe with a decaying
 cosmological term (vacuum energy) that is assumed to be a function of the scale factor.
 In this model, while the cosmological term increases to the early
 universe, the radiation energy density is lower than the model with the cosmological 
"constant". We find that the effects of the decaying cosmological term on the 
expansion rate at the redshift $z<2$ is negligible.

 However, the decrease in the radiation density affects on the thermal history of the 
universe; e.g. the photon decoupling occurs at higher $z$ compared to
the case of the standard $\Lambda$CDM model. 
As a consequence, a decaying cosmological term affects on the cosmic
 microwave background anisotropy. 
We show the angular power spectrum in D$\Lambda$CDM model and compare
with the Wilkinson Microwave Anisotropy Probe (WMAP) data.
}

\FullConference{International Symposium on Nuclear Astrophysics ---
Nuclei in the Cosmos --- IX\\
		 June 25-30 2006\\
		 CERN, Geneva, Switzerland}

\begin{document}

\section{Introduction}

Recent observations (e.g.  type Ia supernovae \cite{SCP2003},
the cosmic microwave background (CMB) \cite{WMAP2006}) indicate that the
cosmological term is necessary. 
If the cosmological term is constant from the Planck time to the
present, there is
the \textit{cosmological constant problem}: the present value of the
cosmological constant is extraordinarily
small compared with an inferred vacuum energy during the Planck time.
To solve this problem, it is natural to consider that the cosmological
term decreases from the large value at the early epoch to the present value. 
Many functional form of the cosmological term has been suggested, 
e.g. the function of the scalar field \cite{Nakamura2006}.
On the other hand, more physically motivated researches of varying vacuum
energy (e.g. cosmic quintessence) have been presented \cite{Lee2006}.

Cosmological constrains on and results of a decaying vacuum energy
density have been investigated, where the ratio of vacuum to radiation
energy was $\sim 4\times10^{-4}$ \cite{Freese1986}.
We note that vacuum energy corresponds to a cosmological term in the
present paper.
The model with a decaying-$\Lambda$ term into the radiation
has been found to affect the thermal evolution of the universe.
Since the radiation temperature is lower compared with the standard
 $\Lambda$CDM (S$\Lambda$CDM) model \cite{Kimura}, 
the molecular formation occurs at earlier
epoch compared to the case of the S$\Lambda$CDM \cite{Kamikawa}.
Furthermore the model is found to be consistent with the CMB temperature
observations at $z<4$ if appropriate parameters are adopted \cite{Puy}.

In the present paper, we assume that  the $\Lambda$ term decays 
into the photon (hereafter we call this the D$\Lambda$CDM model) and investigate the CMB
temperature fluctuation in the D$\Lambda$CDM model. 
Then, we constrain the parameters of the D$\Lambda$CDM model.

\section{A decaying $\Lambda$ cosmology}

Using the Friedmann-Robertson-Walker metric, 
the Einstein equation and the energy-momentum conservation law 
are written as follows:
\begin{equation}
\left( \frac{\dot{a}}{a} \right)^2_{}=
 \frac{8\pi G_N}{3}\rho-\frac{k}{a^2}+\frac{\Lambda}{3},
 \label{eq:friedeq}
\end{equation}
\begin{equation}
 \dot{\rho}+\dot{\frac{\Lambda}{8\pi G^{}_N}}=-3\frac{\dot{a}}{a}\left( \rho+p \right),
\label{eq:rhodot}
\end{equation}
where $a$ is the cosmic scale factor, $k$ is the curvature and
$\Lambda$ is the cosmological term. 
  The total energy density $\rho$ and the pressure are written as
\begin{equation}
 \rho= \rho^{}_m+\rho^{}_\gamma+\rho^{}_{\nu}, \quad 
  p=\frac{1}{3}\left( \rho^{}_{\gamma}+\rho^{}_{\nu}\right),
  \label{eq:rhop}
\end{equation}
where the subscripts $m, \gamma$ and $\nu$ are the non-relativistic matter
(baryon and cold dark matter), photon, neutrino, respectively.
Here the energy density of matter and neutrinos varies as 
$\rho^{}_m=\rho^{}_{m0}a^{-3}_{}$ and $\rho^{}_{\nu
0}=\rho^{}_{\nu}a^{-4}$, 
where the subscript $0$ means the present value.
From eqs. (\ref{eq:rhodot}) and (\ref{eq:rhop}), we get the evolution
of the photon energy density:
\begin{equation}
  \frac{d\Omega^{}_{\gamma}}{da}+4\frac{\Omega^{}_{\gamma}}{a}
	=-\frac{d\Omega^{}_{\Lambda}}{da}.
 \label{eq:emcona}
\end{equation}
with 
\[
\Omega^{}_{i} \equiv \frac{\rho^{}_{i}}{\rho_{\mathrm{crit}}} \ , 
\quad \rho_{\mathrm{crit}} \equiv \frac{3 H_{0}^{2}}{8 \pi G^{}_N} \ ,
\quad \Omega^{}_\Lambda \equiv \frac{\Lambda}{3H_0^2} \ ,
\]
where $H^{}_0$ is the Hubble constant in unit of km/sec/Mpc.

We assume a functional form of $\Lambda$ as follow \cite{Kimura,Kamikawa,Puy,Overduin1998}: 
\begin{equation}
 \Omega^{}_{\Lambda}=\Omega^{}_{\Lambda 1}+\Omega^{}_{\Lambda 2}a^{-m} . \label{eq:lambda}
\end{equation}
Note the present value of $\Omega^{}_{\Lambda}$:
$\Omega^{}_{\Lambda0}=\Omega^{}_{\Lambda 1}+\Omega^{}_{\Lambda 2}$.

\begin{figure}
 \includegraphics[width=.5\linewidth]{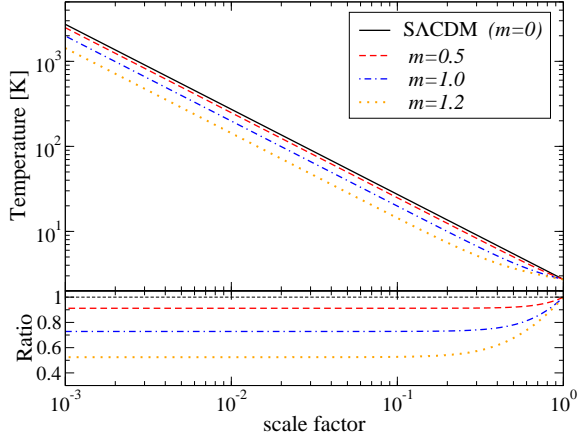}
\caption{Upper panel: the evolution of the photon
 temperature $T_\gamma$ as a function of the scale
  factor in a decaying $\Lambda$ model. Lower panel: the
 ratio of $T_\gamma$ for $m=0.5,~1.0,~1.2$ relative to $m=0$.
}
\label{fig:temp_dlcdm}
\end{figure}

Following the Stefan-Boltzmann's law 
$\rho^{}_{\gamma}\propto T^4_{\gamma}$, 
the photon temperature evolves as follows \cite{Puy}:
\begin{eqnarray}
 T^{}_{\gamma} &=& \frac{T^{}_{\gamma 0}}{a} 
{\left[ 1+\frac{m\Omega_{\Lambda 2}}{\left( 4-m
 \right)\Omega^{}_{\gamma 0}}\left( a^{4-m}_{}-1 \right)
\right]}^{1/4}_{} \texttt{~~for~}m\ne 4 \label{eq:Tgamma_lt4} \\
T^{}_{\gamma} &=& \frac{T^{}_{\gamma 0}}{a}  
\left( 1+4\frac{\Omega^{}_{\Lambda 2}}{\Omega^{}_{\gamma 0}}
\ln{a}\right)^{1/4}_{} \texttt{~~for~} m=4 \label{eq:Tgamma_eq4}
\end{eqnarray}
where 
$\Omega^{}_{\gamma 0}=2.471\times 10^{-5}_{}h^{-2}_{}( T^{}_{\gamma 0}/2.725 {\tt ~K})^4_{}$
is the present photon energy density, 
$h$ is the Hubble constant ($H^{}_{0}\equiv 100 h$ km/sec/Mpc). 
If $m$ and/or $\Omega^{}_{\Lambda 2}$ are too large, 
the photon temperature is negative at some epoch of $a<1$. By excluding
this kind of solution, we obtain limits on
$\Omega^{}_{\Lambda 2}$ and $m$ from eq. (\ref{eq:Tgamma_lt4}): 
$m\Omega^{}_{\Lambda 2}/(4-m)< \Omega^{}_{\gamma 0}$ for $m<4$. 
In this parameter region, 
the first term in eq. (\ref{eq:lambda}) dominate the universe at low-$z$.
As the result, the effects of the second term in
eq. (\ref{eq:lambda}) on the expansion rate is negligible.
For $\Omega^{}_{\Lambda 2}<0$ or $m<0$, we find that $T^{}_{\gamma}$
becomes negative at $a>1$.
Therefore we calculate under the condition of $\Omega^{}_{\Lambda 2}\geq 0$ and $m\geq0$.

Figure \ref{fig:temp_dlcdm} illustrates the evolution of the photon
temperature in the D$\Lambda$CDM model. 
The adopted cosmological parameters are as follows: 
$h=0.73$, $T^{}_{\gamma 0}=2.725$ K, 
$\Omega^{}_{\Lambda 0}=0.763$ and $k=0$~(flat universe).
The photon temperature evolves as
$T^{}_{\gamma}\propto a^{-1}_{}$ and $T^{}_{\gamma}\propto a^{-m/4}_{}$
before and during the $\Lambda$ dominant epoch, respectively. 
Changes in $T^{}_{\gamma}$ affects the cosmic thermal history
significantly \cite{Kamikawa}, which should be constrained by the observations
such as CMB anisotropy as we shall show below.

\section{CMB constraint}

\begin{figure}
 \includegraphics[width=.6\linewidth]{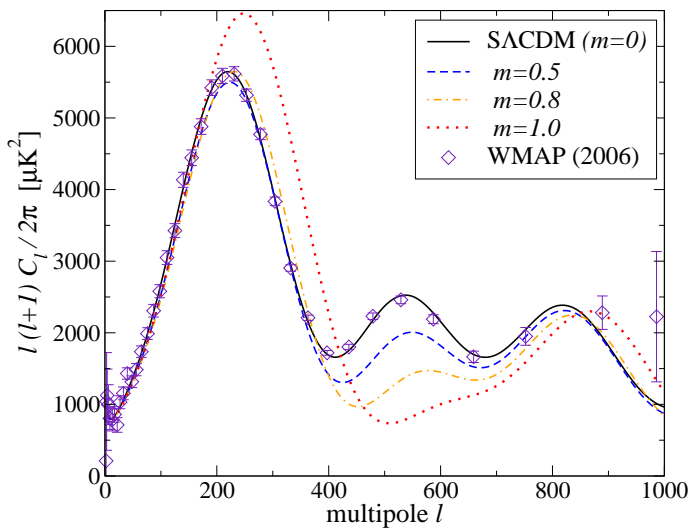}
 \caption{The angular power spectrum in a decaying $\Lambda$ cosmology
 and WMAP observation data \cite{WMAP2006}. The solid line is the result for the S$\Lambda$CDM
 model. The dashed, the dot-dashed and the dotted lines are those of D$\Lambda$CDM 
model with $(\Omega,^{}_{\Lambda 2},m)=(10^{-4}, 0.5)$, $(10^{-4}, 1.0)$ and
 $(10^{-4}, 1.2)$, respectively.}
\label{fig:cl_dlcdm}
\end{figure}

The CMB anisotropy observed by the WMAP constrains the cosmological model
with very high accuracy. In this section, 
we investigate the consistency of D$\Lambda$CDM model with WMAP and give
the limit to the model parameters.

We calculate the CMB power spectrum by 
modifying CMBFAST code \cite{cmbfast}.
Figure \ref{fig:cl_dlcdm} shows the angular power spectrum in the
D$\Lambda$CDM model. We adopt the following cosmological parameters:
the baryon density parameter $\Omega^{}_bh^2_{}=0.0223$ and the CDM
density parameter $\Omega^{}_{CDM}h^2_{}=0.104$. 
We neglect reionization. 
If $m$ (and/or $\Omega^{}_{\Lambda 2}$) is small, the amplitude of the power
spectrum decreases. If we take larger values of $m$, 
the first and third peaks of the CMB power spectrum increases
because of the large baryon density relative to
the photon energy density. Furthermore the CMB power spectrum shifts toward
higher-$l$ because the photon last scattering occurs at an earlier epoch.

\begin{figure}
 \includegraphics[width=0.5\linewidth]{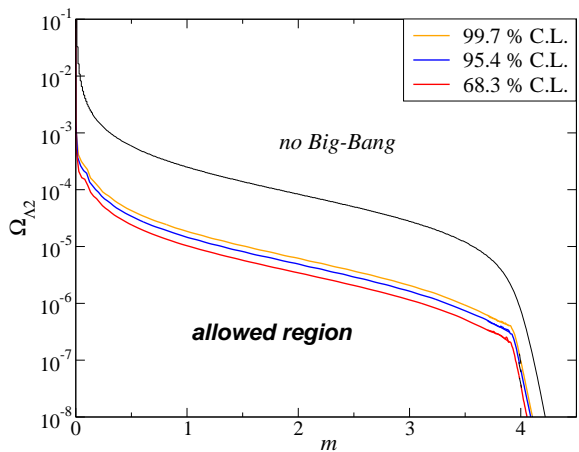}
\caption{Constraint on $m-\Omega^{}_{\Lambda 2}$ plane from WMAP.
Drawn lines correspond to 1, 2 and 3$\sigma$ confidence limit.
The labeled {\it no Big-Bang} region means that 
$T^{}_{\gamma}$ is negative at $a<1$.}
\label{fig:contour}
\end{figure}

To obtain the upper limit of $\Omega^{}_{\Lambda 2}$ and $m$,
we calculate the likelihood function given by Ref. \cite{Verde2003}. 
Figure \ref{fig:contour} shows 68.3\%, 95.4\% and 99.7\% confidence
limits on the $m-\Omega^{}_{\Lambda 2}$ plane from CMB.
We obtain the constraint 
$m\Omega^{}_{\Lambda 2}/\left( 4-m \right) < 4.9\times 10^{-6}_{}$ at
95 \% confidence limit. 
With the value of this upper limit, the photon last scattering occurs at
the earlier epoch 
by $\Delta z\sim 30$ compared with that in the S$\Lambda$CDM model. 
Our constraint is severer than that from the observed radiation
temperature : 
$|m| \le 1, |\Omega^{}_{\Lambda,2}|\le 10^{-4}_{}$ \cite{Puy}.
Therefore, the  results indicate that a decaying-$\Lambda$ contribution to the cosmic thermal
evolution should be small.

It should be noted that $\Omega_bh^2$ has been fixed during the
calculation for simplicity. If we operate CMBFAST with $\Omega_b$
varying, we can get the more reasonable parameter regions 
for the D$\Lambda$CDM model. 
Then the primordial abundances of He, D and Li would be different from those
predicted by WMAP.

\begin{thebibliography}{99}
 \bibitem{SCP2003}
	J.~L.~Tonry {\it et al.}  [Supernova Search Team Collaboration],
	\emph{Cosmological Results from High-z Supernovae}
	, \emph{Astrophys. J.}  {\bf 594} (2003) 1  [{\tt arXiv:astro-ph/0305008}].
\bibitem{WMAP2006}
	D.~N.~Spergel {\it et al.}, 
	\emph{Wilkinson Microwave Anisotropy Probe (WMAP) three year results:
	Implications for cosmology} {\tt arXiv:astro-ph/0603449};
	G.~Hinshaw {\it et al.}
	, \emph{Three-year Wilkinson Microwave Anisotropy Probe (WMAP)
	observations: Temperature analysis}, {\tt arXiv:astro-ph/0603451}.
 \bibitem{Nakamura2006}
	R.~Nakamura, M.~Hashimoto, S.~Gamow and K.~Arai,
	\emph{Big-bang nucleosynthesis in Brans-Dicke cosmology with a varying Lambda
	term related to WMAP}, \emph{Astron.\ Astrophys.}  {\bf 448} (2006) 23
	[{\tt arXiv:astro-ph/0509076}].
\bibitem{Lee2006}
	 S.~Lee, G.~C.~Liu and K.~W.~Ng,
	 \emph{Constraints on the coupled quintessence from cosmic microwave  background
	 anisotropy and matter power spectrum},
	 \emph{Phys.\ Rev.\ D} {\bf 73} (2006) 083516.
	 [{\tt arXiv:astro-ph/0601333}].
\bibitem{Freese1986}
	 K.~Freese, F.~C.~Adams, J.~A.~Frieman and E.~Mottola,
	 \emph{Cosmology With Decaying Vacuum Energy},
	 \emph{Nucl.  Phys. B} {\bf 287}, (1987) 797.
\bibitem{Kimura} 
	K. Kimura, M. Hashimoto, M. Sakoda \& K. Arai
	, \emph{Effects on the Temperatures of a Variable Cosmological Term after Recombination}
	, \emph{Astropphys. J. Letter} {\bf 561}, (2001) L19 
\bibitem{Kamikawa}
	M. Hashimoto, T. Kamikawa \& K. Arai
	, \emph{Effects of a Decaying Cosmological Term on the Formation of Molecules and First Objects}
	, \emph{Astrophys. J.} {\bf 598} (2003) 13 
\bibitem{Puy}
	D. Puy, \emph{Thermal balance in decaying $\Lambda$
	cosmologies},
	\emph{Astron. \& Astrophys.} {\bf 422} (2004) 1 
 \bibitem{Overduin1998}
	J.~M.~Overduin and F.~I.~Cooperstock,
	\emph{Evolution of the Scale Factor with a Variable Cosmological Term}
	, \emph{Phys. Rev. D} {\bf 58} (1998) 043506 [{\tt arXiv:astro-ph/9805260}].
\bibitem{cmbfast}
	U.~Seljak and M.~Zaldarriaga,
	\emph{A Line of Sight Approach to Cosmic Microwave Background Anisotropies},
	\emph{Astrophys. J.} {\bf 469} (1996) 437 [{\tt arXiv:astro-ph/9603033}].
\bibitem{Verde2003}
	L.~Verde {\it et al.},
	\emph{First Year Wilkinson Microwave Anisotropy Probe (WMAP)
	Observations: Parameter Estimation Methodology}
	, \emph{Astrophys. J. Suppl.}  {\bf 148} (2003) 195
	[{\tt arXiv:astro-ph/0302218}].
\end{thebibliography}
\end{document}